\documentclass[traditabstract]{aa}

\usepackage{graphicx}
\usepackage{enumerate}
\usepackage{natbib}
\usepackage{txfonts}
\usepackage{verbatim}
\bibpunct{(}{)}{;}{a}{}{,}    % to follow the A&A style

\begin{document}

%%%%%%%%%%%%%%%%%%%%%%%%%%%%%%%%%%%%%%%%%%%%%%%%%%%%%%%%%%%%%%%%%%%%%%%%
\title{\mbox{3-D} non-LTE radiative transfer effects in \ion{Fe}{i} lines}
\subtitle{III. Line formation in magneto-hydrodynamic atmospheres}
%%%%%%%%%%%%%%%%%%%%%%%%%%%%%%%%%%%%%%%%%%%%%%%%%%%%%%%%%%%%%%%%%%%%%%%%

\author{R.~Holzreuter\inst{1,2}, S.~K.~Solanki\inst{1,3} }

\institute{MPI for solar system research,  Katlenburg-Lindau, Germany \and 
  Institute of Astronomy, ETH Zentrum, CH-8092 Zurich, Switzerland \and 
  School of Space Research, Kyung Hee University, Yongin, Gyeonggi 446-701, Republic of Korea\\
  \email{holzreuter@astro.phys.ethz.ch}}

\offprints{R.~Holzreuter}

\date{Received $<$date$>$; accepted $<$date$>$}

%%%%%%%%%%%%%%%%%%%%%%%%%%%%%%%%%%%%%%%%%%%%%%%%%%%%%%%%%%%%%%%%%%%%%%%%
%%%%%%%%%%%%%%%%%%%%%%%%%%%%%%%%%%%%%%%%%%%%%%%%%%%%%%%%%%%%%%%%%%%%%%%%
\abstract{
Non-local thermodynamic equilibrium (NLTE) effects in diagnostically important solar \ion{Fe}{i} lines are important due to the strong sensitivity of \ion{Fe}{i} to ionizing UV radiation, which may lead to a considerable under-population of the \ion{Fe}{i} levels in the solar atmosphere and, therefore, to a sizeable weakening of \ion{Fe}{i} lines. Such NLTE effects may be intensified or weakened by horizontal radiative transfer (RT) in a three-dimensionally (\mbox{3-D}) structured atmosphere. We analyze the influence of horizontal RT on commonly used \ion{Fe}{i} lines in a snapshot of a \mbox{3-D} radiation magneto-hydrodynamic (MHD) simulation of a plage region. NLTE- and horizontal RT effects occur with considerable strength (up to 50\% in line depth or equivalent width) in the analyzed snapshot. As they may have either sign and both signs occur with approximately the same frequency and strength, the net effects are  small when considering spatially averaged quantities. The situation in the plage atmosphere turns out to be rather complex. Horizontal transfer leads to line-weakening relative to \mbox{1-D} NLTE transfer near the boundaries of $kG$ magnetic elements. Around the centers of these elements, however, we find an often significant line-strengthening. This behavior is in contrast to that expected from previous \mbox{3-D} RT computations in idealized flux-tube models, which display only a line weakening.   The origin of this unexpected behavior lies in the fact that magnetic elements are surrounded by dense and relatively cool down-flowing gas, which forms the walls of the magnetic elements. The continuum in these dense walls is often formed in colder gas than in the central part of the magnetic elements. Consequently, the central parts of the magnetic element experience a sub-average UV-irradiation leading to the observed \mbox{3-D} NLTE line strengthening.
}

\maketitle

%%%%%%%%%%%%%%%%%%%%%%%%%%%%%%%%%%%%%%%%%%%%%%%%%%%%%%%%%%%%%%%%%%%%%%%%
\section{Introduction}\label{sec:fluxsheet_intro}
%%%%%%%%%%%%%%%%%%%%%%%%%%%%%%%%%%%%%%%%%%%%%%%%%%%%%%%%%%%%%%%%%%%%%%%%
The Sun's atmosphere is strongly structured at all spatial scales that can be resolved. In particular, the magnetic field in the Sun's photospheric layers is concentrated at scales that have only begun to be resolved recently \citep[e.g.][]{laggetal2010,martinezgonzalezetal2012}. In the face of limited spatial resolution, spectral lines computed in 
\mbox{3-D} model atmospheres resulting from radiation MHD simulations may be used to probe properties of magnetic features at a spatial resolution exceeding that of the best observations available today. However, the task is a difficult one and therefore simplifying assumptions have to be made. In the past, an often used simplification was the application of local thermodynamic equilibrium (LTE) to the calculation of atomic level populations. By doing so, the influence of the radiation field on the level populations and, therefore, on the line formation, was completely neglected. An important step towards a higher degree of realism in the calculation of population numbers was the introduction of non-LTE (NLTE) calculations in one-dimensional (\mbox{1-D}) atmospheric models, still neglecting the influence of horizontal radiative transfer (RT) \citep[e.g.,][]{athaylites1972, steenbockholweger1984, boyarchuketal1985, solankisteenbock1988, theveninidiart1999, shchukinatrujillobueno2001}. The inclusion of proper \mbox{3-D} NLTE RT reflects the next step in improving the realism of line profile calculations.

In \citet{holzreutersolanki2012} (hereafter referred to as \mbox{Paper I}), we investigated the effects of \mbox{3-D} NLTE RT in several idealized (i.e. thin-tube) models of flux sheets (FS) and flux tubes (FT), qualitatively confirming the results of \citet{stenholmstenflo1977,stenholmstenflo1978} who found that the line profiles of iron lines in a simplified FT may be strongly weakened by horizontal RT \citep[cf.][]{brulsvdluehe2001}. The strong UV irradiation from the surrounding hot walls of the FT represented the main factor producing this line weakening by the so-called UV over-ionization \citep[see, e.g.,][ and references therein]{athaylites1972, rutten1988, shchukinatrujillobueno2001}. However, the true effect was found to be much smaller than originally calculated by \citet{stenholmstenflo1977} because the UV opacity used in their calculation was too low as they neglected the opacity contributions from the uncounted iron absorption lines in the UV wavelength range. 

A natural next step towards increased realism was taken by \citet{holzreutersolanki2013} (hereafter called \mbox{Paper II}), where we used a snapshot of a \mbox{3-D} radiation hydrodynamic (HD) simulation computed with the MURAM code \citep{voegleretal2005} as the atmospheric input, thus extending the work of \citet{KiselmanNordlund1995}. In such an atmosphere, the line weakening effect by \mbox{3-D} NLTE RT due to hot UV irradiation, acting mainly in intergranular lanes, occurs alongside the opposite effect of line strengthening in regions with colder surroundings, i.e. mostly at the centers of granules. Such a strengthening was found already in \mbox{Paper I} at a few positions, but with a negligible influence on the spatially averaged line profile. The much stronger and more frequent occurrence in the HD atmosphere leads to very similar \emph{average} profiles of the \mbox{1-D} and \mbox{\mbox{3-D}} NLTE calculations. The spatially \emph{resolved} line profiles, however, may differ considerably between the three calculation methods (LTE, \mbox{1-D} NLTE, and, \mbox{3-D} NLTE) with corresponding implications for the interpretation and inversion of high-resolution observations. 

Here, we continue our investigation on the effects of true \mbox{3-D} RT on diagnostically important iron lines by considering a snapshot of a magneto-hydrodynamic (MHD) simulation of a plage region calculated by the MURAM code. The introduction of a (strong) magnetic field in the MHD simulation leads to the formation of evacuated magnetic elements. We show that the heavily idealized models of \citet{stenholmstenflo1977, stenholmstenflo1978, brulsvdluehe2001} and \mbox{Paper I} can only partially reproduce the far more complex situation of magnetic elements in an MHD atmosphere. Note that the \mbox{3-D} formation of chromospheric spectral lines in \mbox{3-D} radiation MHD simulation has been investigated by \citet{leenaartsetal2009,leenaartsetal2012,leenaartsetal2013} and \citet{leenaarts2010}

%%%%%%%%%%%%%%%%%%%%%%%%%%%%%%%%%%%%%%%%%%%%%%%%%%%%%%%%%%%%%%%%%%%%%%%%
\section{Model ingredients}\label{sec:fluxsheet_model}
%%%%%%%%%%%%%%%%%%%%%%%%%%%%%%%%%%%%%%%%%%%%%%%%%%%%%%%%%%%%%%%%%%%%%%%%

\subsection{Radiative transfer calculations}\label{sec:mhdfe23_model_RT}
%%%%%%%%%%%%%%%%%%%%%%%%%%%%%%%%%%%%%%%%%%%%%%%%%%%%%%%%%%%%%%%%%%%%%%%%
All calculations were done in the same way as in \mbox{Paper II}, except that the full Stokes vector of the Zeeman split lines was computed. For a detailed description, especially of the formal solution of the full Stokes vector problem, we refer to \mbox{Paper I} (Appendix I and II). Here, we present only a short summary.

We used the RH code \citep{uitenbroek2000} with adaptations as mentioned in \mbox{Papers I and II}, the most important being the interpolation method of the formal solution along the short characteristics, where we used monotonic parabolic B\'ezier integration as proposed by \citet{auer2003} together with an enhancement we introduced for the magnetic case (see the Appendix B of \mbox{Paper I}). 

The atomic model was the same as used in \mbox{Paper II} (23 levels, 33 lines). It was tailored to optimally reproduce the NLTE effects on the population numbers of levels involved in the formation of the transitions investigated here, i.e. the \ion{Fe}{i} $524.71$~nm/$525.02$~nm and the $630.15$~nm / $630.25$~nm line pairs, and, at the same time, to be as small as possible to reduce the computational effort. The problem of strong UV over-ionization owing to missing UV opacity \citep{brulsetal1992} was --- again --- eliminated by artificially increasing the opacities in the relevant wavelength range (opacity fudging) as proposed by \citet{brulsetal1992}. Refer to \mbox{Paper I} for a more complete discussion.

The full Stokes vector was calculated in the same way as in \mbox{Paper I}, i.e. by using the so-called field-free method of \citet{rees1969}. Thus, the influence of the Zeeman effect on the line profiles (splitting, polarization) was included only in the final determination of the emerging intensity but not during the iteration of the level population numbers. In the following, however, we analyze only the Stokes $I$ profiles since we are interested here mainly in uncovering the basic effects. The influence of neglecting \mbox{3-D} NLTE on inversions of Stokes profiles will be considered in a following paper.

\subsection{Model atmospheres}\label{sec:mhdfe23_model_atmos}
%%%%%%%%%%%%%%%%%%%%%%%%%%%%%%%%%%%%%%%%%%%%%%%%%%%%%%%%%%%%%%%%%%%%%%%%
\begin{figure*}
\center{ \resizebox{0.76\width}{!}{\includegraphics{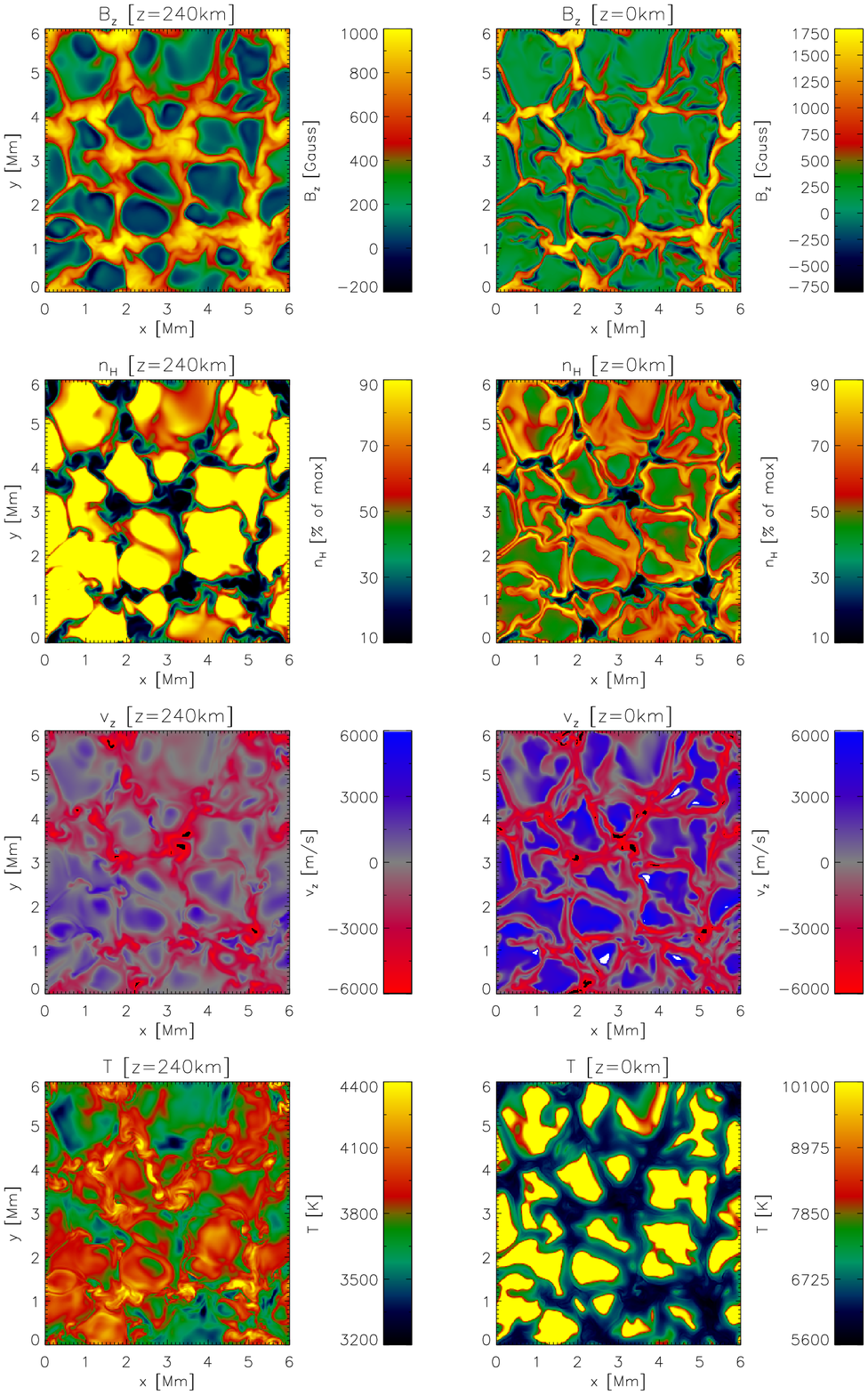}} }
 \caption{ 
 Atmospheric parameters as indicated above each panel at heights $z=0$~km (right panels) corresponding to $<\tau_c>=1$, and, $z=240$~km, the approximate average line core formation region of the investigated lines (left panels). Positive velocities ($v_z>0$~ km) correspond to up-flows (blue).
} 
 \label{fig:tv_overview_atmos}
\end{figure*}

The input model for our calculations was taken from a realistic \mbox{3-D} radiation magneto-hydrodynamic simulation of a plage region. We used a snapshot from a $B=400$ G run of the MURAM code \citep{voegleretal2005}, i.e., a run with an initial unipolar (positive polarity) vertical homogeneous magnetic field that was imposed on a relaxed HD run. The original resolution of the cube was $288 \times 288 \times 120$ voxels spanning a geometric range of approximately $6$~Mm$~\times~6$~Mm$~\times~1.6$~Mm. 

The construction of the atmospheres (\mbox{3-D} and \mbox{1-D}) was performed in the same way as in \mbox{Paper II}. However, only the bottom 30 grid points were removed instead of 48 as in the HD case of \mbox{Paper II}) owing to the strong fields in the present atmosphere and the associated evacuation, which pushes the\ $log\ \tau =0$ surface to deeper layers. The resulting  $288 \times 288 \times 90$ cube starts approximately $500$~km below the average $\tau_c^{500}=1$ level ($z=0$~km) and reaches a height of approximately $750$~km above it. The minimum $\tau_c^{500}$ value at the bottom of the reduced cube is approximately $17$, while the average value at that depth amounts to above $1000$. 

Maps of the most important physical parameters of the input atmosphere are given in Fig.~\ref{fig:tv_overview_atmos} for the same two geometric heights as in \mbox{Paper II}, i.e. at $z=0$~km and $z=240$~km, the latter very roughly being the height of formation of the cores of the lines used in this work. Note that the granular-intergranular temperature inversion also takes place at these heights. Some parts --- mostly in the lower half of the image --- still show hot uprising granules, whereas those in the upper part consist of hotter intergranular lanes. The panel in Fig.~\ref{fig:tv_overview_atmos} showing the vertical velocity ($v_z$) confirms this view, as the granules in the lower part still show considerable upward velocities (blue), whereas those in the upper part of the image barely host rising gas at all (grey). Note also, that the downward velocities are often higher at the boundaries of the magnetic elements, whereas the velocities in the centers of the magnetic elements are often reduced owing to the magnetic field, although at some locations very strong downflows are found inside the magnetic features. 

Maps of the vertical component of the magnetic field, $B_z$ (top panels of Fig.~\ref{fig:tv_overview_atmos}), reveal field concentrations along intergranular lanes (in the form of strongly elongated magnetic elements similar to flux sheets) and at the intersections of intergranular lanes (similar to flux tubes). For simplicity in the following we refer to the elongated magnetic elements as flux sheets, or FS, and to the rounder ones as flux tubes, or FT. Only few areas are found with negative vertical fields and the field strength there is much lower than that in the magnetic elements. They are mostly located at the boundaries between intergranular and granular areas where the down-flows are strongest \citep[see][]{buehleretal2015,buehler2014}.

%%%%%%%%%%%%%%%%%%%%%%%%%%%%%%%%%%%%%%%%%%%%%%%%%%%%%%%%%%%%%%%%%%%%%%%%
\section{Results}\label{sec:mhdfe23_results}
%%%%%%%%%%%%%%%%%%%%%%%%%%%%%%%%%%%%%%%%%%%%%%%%%%%%%%%%%%%%%%%%%%%%%%%%

\subsection{Continuum intensity}\label{sec:mhdfe23_results_continuum}
%%%%%%%%%%%%%%%%%%%%%%%%%%%%%%%%%%%%%%%%%%%%%%%%%%%%%%%%%%%%%%%%%%%%%%%%
\begin{figure}
\center{ \resizebox{0.75\width}{!}{\includegraphics{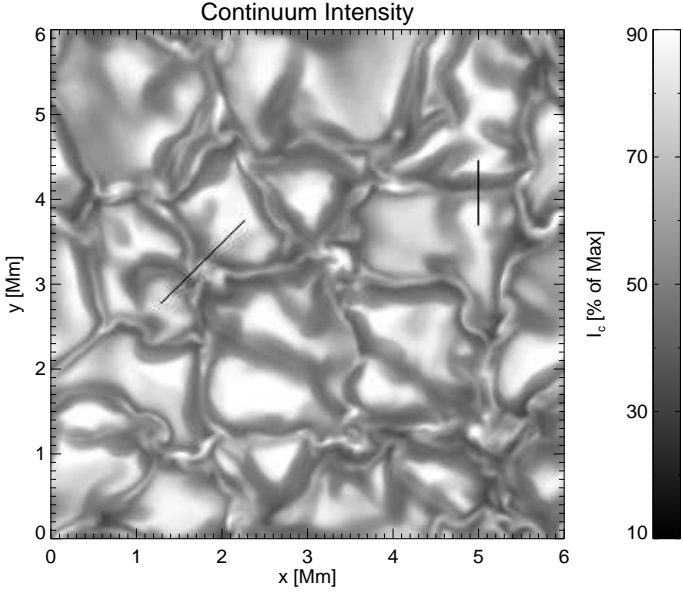}} }
 \caption{ 
 Continuum intensity at $\approx 525$~nm. The grey scale has been clipped to 10\% and 90\%, respectively, of the maximum continuum intensity value. The two black lines indicate the selected cuts through two distinct magnetic elements, namely, an elongated feature similar to a flux sheet (FS) and a more compact one that we will call a flux tube (FT).
} 
 \label{fig:tv_overview_Icont}
\end{figure}

The computed intensity at a continuum wavelength close to $525$~nm is presented in Fig.~\ref{fig:tv_overview_Icont}. Two types of regions of more intense radiation can be found: One in the hot granules and one in the cores of magnetic flux concentrations. Lower intensities are present in the intergranular lanes between magnetic FS and granules. %The resulting RMS contrast for the continuum near $630$~nm is $17.3$\%, that close to $525$~nm amounts to $23.1$\%. 
The results of our three calculation methods (\mbox{3-D} NLTE, \mbox{1-D} NLTE, LTE) show no differences in the continuum intensity as it is formed strictly in LTE in our computation. Note that we neglect the possible departures from LTE found by \citet{shapiroetal2010}. The two black lines indicate cuts through an FT and an FS, respectively, which are investigated in detail in later sections.

\subsection{Influence of NLTE and horizontal RT}\label{sec:mhdfe23_results_method}
%%%%%%%%%%%%%%%%%%%%%%%%%%%%%%%%%%%%%%%%%%%%%%%%%%%%%%%%%%%%%%%%%%%%%%%%
In this section, we compare the results of LTE, \mbox{1-D} NLTE and \mbox{3-D} NLTE radiative transfer on the spectral lines. We are looking for the effects of horizontal RT as described by \citet{stenholmstenflo1977,stenholmstenflo1978}, who found a pronounced weakening of the lines calculated in \mbox{3-D} NLTE in magnetic elements. Such a weakening may lead to errors when diagnosing atmospheric parameters such as the temperature.  To simplify the discussion, we introduce the following notation for the relative differences of intensities (I) of the different calculation methods (\mbox{3-D} NLTE, \mbox{1-D} NLTE and LTE):

\begin{equation}\label{eq:abbrev_ew}
\delta I^{M_i-M_j}
\,\,=\,\,
\frac{ I^{M_i}-I^{M_j} } { I^{M_j} } 
\rm \ ,
\end{equation}

with each $M_i$, and $M_j$ referring to one out of the three calculation methods (3D, 1D, LTE), and $I^{M_i}$ being the lowest intensity value of the line obtained with method $M_i$ at a specific spatial position. In the following, we will call $I^{M_i}$ residual intensity of method $M_i$. Analogously, for the equivalent widths (EW) we define

\begin{equation}\label{eq:abbrev_ew}
\delta E^{M_i-M_j}
\,\,=\,\,
\frac{ EW^{M_i}-EW^{M_j} } { EW^{M_j} } 
\rm \ .
\end{equation}

Note that $\delta E^{1D-3D}>0$ and $\delta I^{1D-3D}<0$ if the \mbox{3-D} NLTE line is weaker than its \mbox{1-D} NLTE counterpart, i.e., $\delta E^{1D-3D}$ is a measure for the line weakening of the \mbox{3-D} NLTE line when compared with the \mbox{1-D} NLTE line strength. The larger the $\delta E^{1D-3D}$ is, the stronger is the line weakening. Negative $\delta E^{1D-3D}$ denote locations with line strengthening.

\subsubsection{Equivalent width}\label{sec:mhdfe23_results_method_EW}
%%%%%%%%%%%%%%%%%%%%%%%%%%%%%%%%%%%%%%%%%%%%%%%%%%%%%%%%%%%%%%%%%%%%%%%%
\begin{figure*}
\center{ \resizebox{0.85\width}{!}{\includegraphics{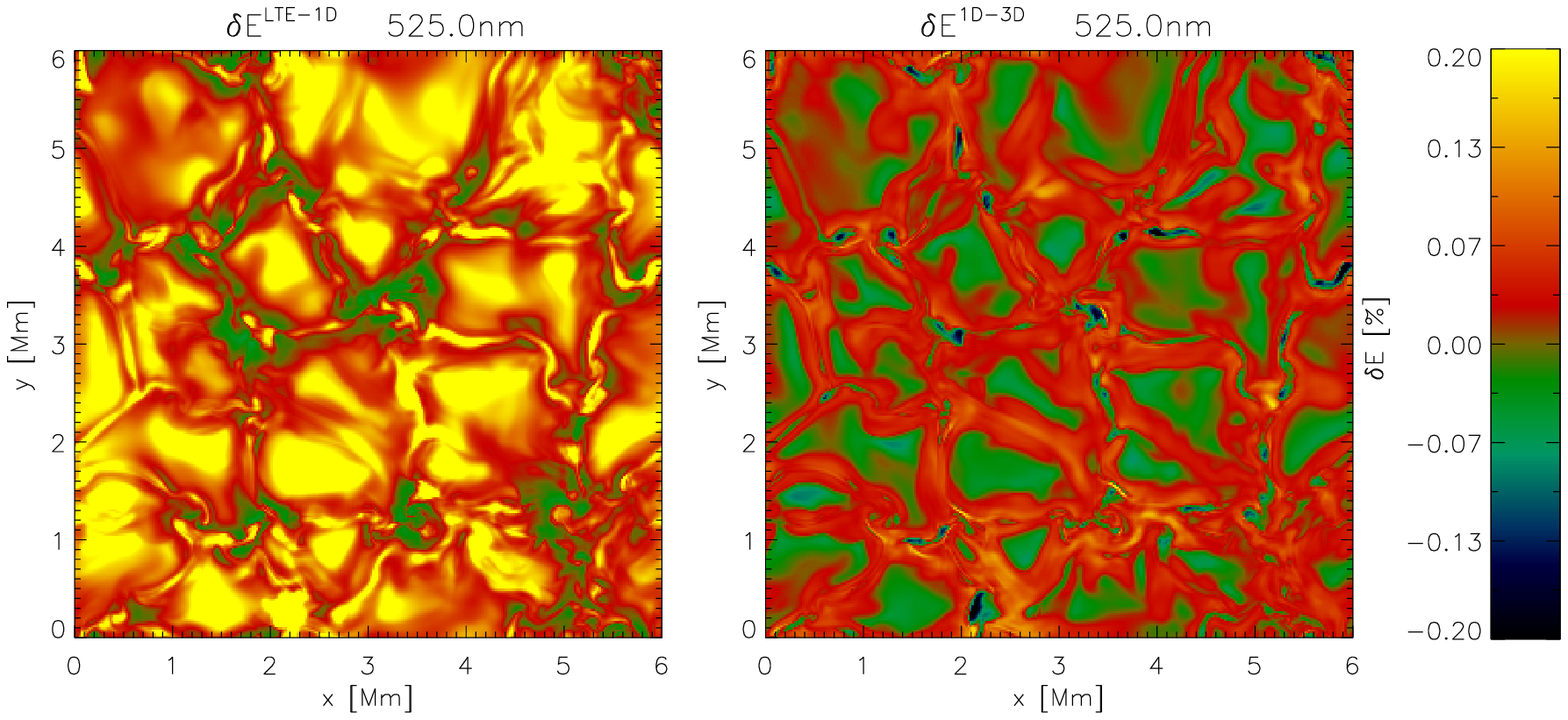}} }
\center{ \resizebox{0.85\width}{!}{\includegraphics{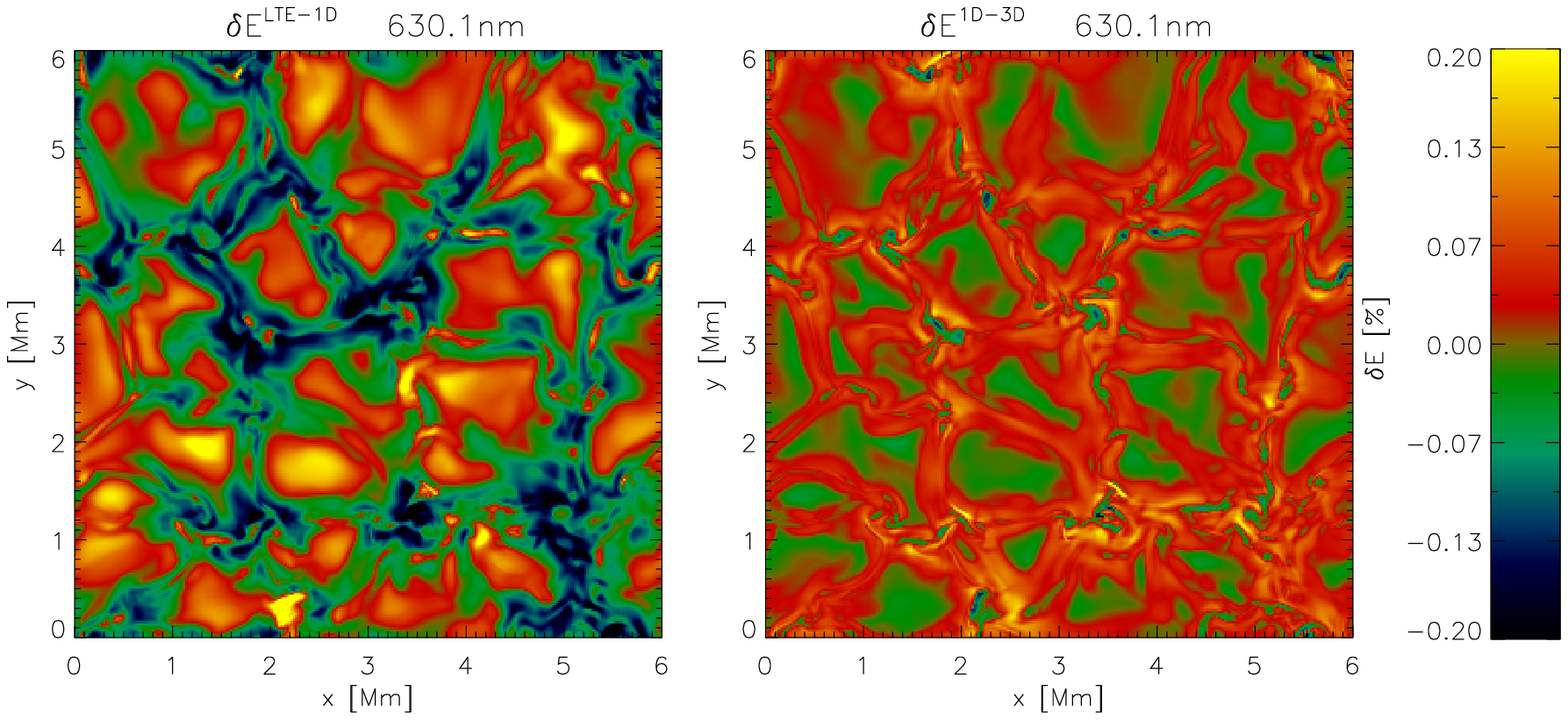}} }
 \caption{ 
 Spatial distribution of $\delta E^{LTE-1D}$ (left panels) and $\delta E^{1D-3D}$ (right panels) for the $525.02$~nm (upper panels) and the $630.15$~nm line (lower panels). Positive $\delta E^{1D-3D}$ indicates locations with weaker \mbox{3-D} NLTE than the corresponding \mbox{1-D} lines.
} 
 \label{fig:tv_ratio_Aeqw}
\end{figure*}

Figure  \ref{fig:tv_ratio_Aeqw}  shows the spatial distribution of $\delta E^{LTE-1D}$ (left panels) and $\delta E^{1D-3D}$ (right panels). Upper panels: $525.02$~nm; lower panels: $630.15$~nm. Yellow and red colors indicate areas where the \mbox{1-D} NLTE (\mbox{3-D} NLTE) line is weaker than the corresponding LTE (\mbox{1-D} NLTE) line. The results for the $524.71$~nm and the $630.25$~nm lines are not shown, since they agree qualitatively with those of the other line in the respective line pair. The effects in the $630.15$~nm line are slightly weaker than those found in the $630.25$~nm line.

Looking at Fig.~\ref{fig:tv_ratio_Aeqw}, we find large values in all panels. The influence of NLTE (left panels) is typically larger than that of horizontal RT (right panels) and often has the reversed sign, i.e., NLTE effects are partially cancelled by those of horizontal RT if one compares \mbox{3-D} NLTE with LTE calculations.

Where are these differences located spatially? First consider the distribution of $\delta E^{LTE-1D}$ in the left panels of Fig.~\ref{fig:tv_ratio_Aeqw}. In the granules and in the centers of the magnetic elements we find $\delta E^{LTE-1D}>0$, whereas $\delta E^{LTE-1D}<0$ in the periphery of the magnetic elements. The positive and negative deviations  approximately reach the same strength (up to $20$\%) for the \ion{Fe}{i} $630.15$~nm line. For the $525.02$~nm line, $\delta E^{LTE-1D}$ is very large in the granules and the centers of the magnetic elements and close to zero elsewhere.

This behavior results from the fact that the relatively weak $525.02$~nm line's source function (LTE and NLTE) mainly follows the local Planck function. At locations with a steep vertical temperature gradient (e.g. at a granule), UV irradiation from below ionizes iron and thus reduces the population of the lower level. The resulting lower opacity of the NLTE line shifts its formation height towards deeper and hotter regions in the atmosphere, leading to an increase of the source function and therefore a weakening of the line in NLTE. In regions with smaller temperature gradient, the effect of NLTE disappears almost completely. These arguments apply to the \ion{Fe}{i} $630.15$~nm line as well. However, for the \ion{Fe}{i} $630.15$~nm line an additional effect is found that acts with a reversed sign. The $630.15$~nm line is much stronger than the $525.02$~nm line. Therefore, it is formed higher and the source function falls clearly below the Planck function, making the NLTE line generally stronger. This balance between the two effects (over-ionization and low source function) leads to the pattern seen in the lower left panel of Fig.~\ref{fig:tv_ratio_Aeqw}, with the line being stronger in LTE in the granules, but weaker in the intergranular lanes.

The right panels of Fig.~\ref{fig:tv_ratio_Aeqw} display the influence of horizontal RT. Both lines exhibit almost the same pattern, the line weakening in \mbox{3-D} NLTE being slightly more pronounced for the \ion{Fe}{i} $630.15$~nm line. In the magnetic elements, with the exception of their central parts, the \mbox{3-D} NLTE lines are weaker (red and yellow parts in the right panels of Fig.~\ref{fig:tv_ratio_Aeqw}), as expected from the investigation of \citet{stenholmstenflo1977} and \mbox{Paper I}, owing to the strong irradiation from the hot environment. 

However, the situation in the central parts of the magnetic elements is reversed (small blue and green patches within the red and yellow lanes in the right panels of Fig.~\ref{fig:tv_ratio_Aeqw}): At these locations, we find the \mbox{3-D} NLTE lines to be clearly stronger. This is a surprising result that contradicts the result from the idealized models of \citet{stenholmstenflo1977, brulsvdluehe2001} and \mbox{Paper I}, where the line weakening effect of horizontal transfer was strongest in the centers of the elements. Obviously, the idealized models differ significantly from the more realistic picture provided by the MHD simulations.

In large parts of the granules $\delta E^{1D-3D}$ is negative, a result already found in \mbox{Paper II}. The origin of this effect lies in the influence of the cooler horizontal environment near the line forming region, i.e. the intergranular lanes.

\begin{figure*}
\center{ \resizebox{0.85\width}{!}{\includegraphics{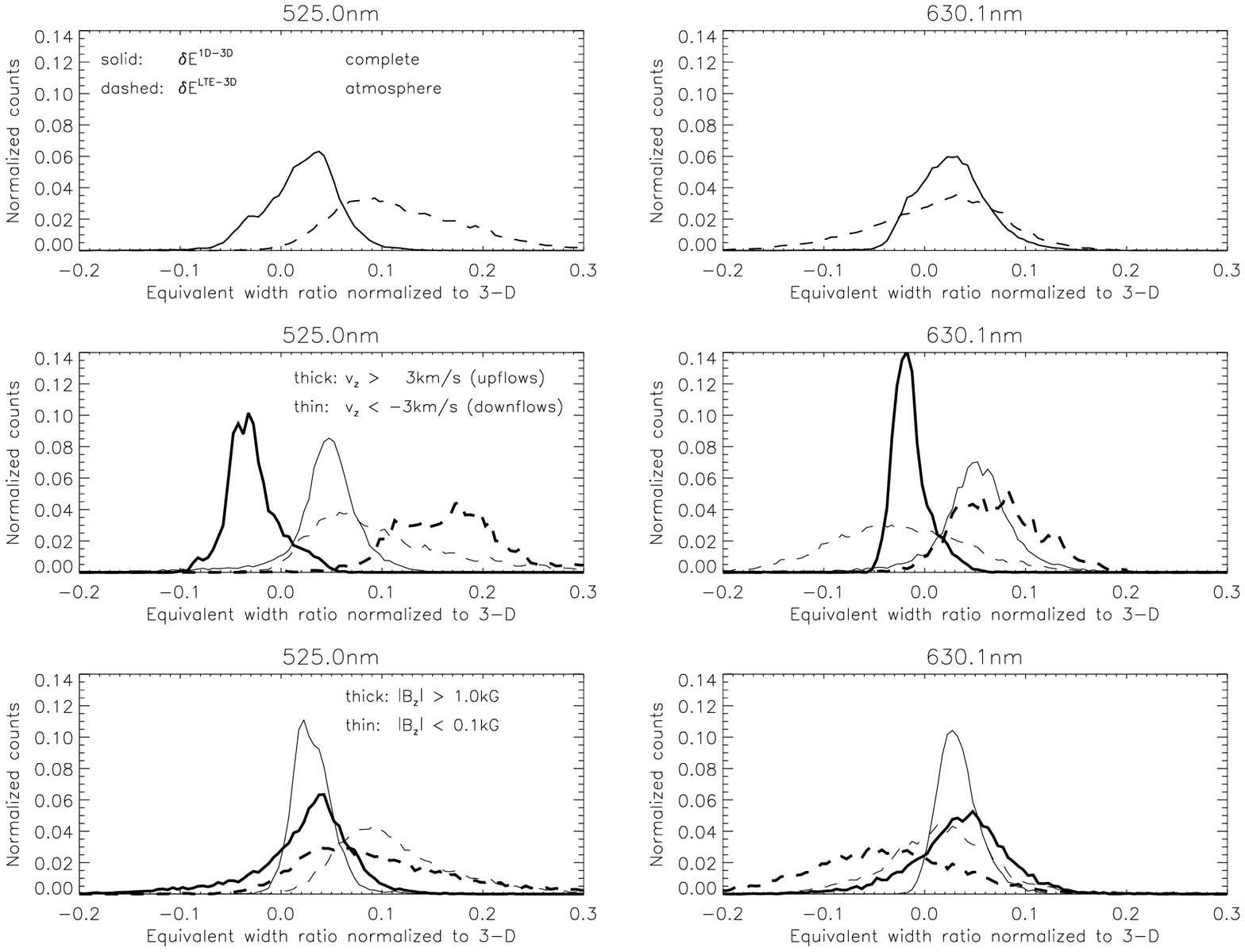}} }
 \caption{ 
 Distribution of the $\delta E^{LTE-3D}$ (dashed lines) and $\delta E^{1D-3D}$ (solid lines) values for the $525.02$~nm (left panels) and the $630.15$~nm line (right panels). The three vertically arranged panels contain the histograms for different spatial selections. Top panels: full atmosphere; Middle panels: Areas with strong up-flows (thin lines) and down-flows (thick lines), the spatial points were selected if the vertical velocity at $\tau_c=1$ was larger than $3$~km/s in either direction. Bottom panels: The same as in the middle panels, but with the thick line now representing points at which the magnetic field strength $|B_z|$ is larger than $1$ kG, and the thin line field strengths below $100$~G. The number of selected spatial points in the middle and bottom panels amounts to approximately $10$\% to $20$\% of the whole atmosphere for each selection.
} 
 \label{fig:EWratio_regions}
\end{figure*}

Figure \ref{fig:EWratio_regions} depicts histograms of the weakening/strengthening of the $525.02$~nm  and the $630.15$~nm lines calculated in LTE and \mbox{1-D} NLTE in relation to the \mbox{3-D} NLTE line profile. While the top panels present the distributions for the whole snapshot, the middle and lower panels shows histograms for selected subsets of the atmosphere. In the middle panels, the selection represents areas of strong up- (thin) and down-flows (thick), respectively. The selected points all have $|v_z|>3$~km/s at the  $\tau_c = 1$ level. As expected from Fig.~\ref{fig:tv_ratio_Aeqw}, we find $\delta E^{1D-3D}>0$ in the down-flow areas and $\delta E^{1D-3D}<0$ in the  up-flows areas and the reversed situation for the $\delta E^{LTE-3D}$ because the LTE - \mbox{1-D} NLTE difference is stronger than that between \mbox{1-D} and \mbox{3-D} NLTE. The width of the $\delta E^{LTE-3D}$ distribution is often quite large, especially for the $525.02$~nm line.

In the bottom panels we distinguish between the subset of pixels with $|B_z|>1$~kG (called "magnetic") and those with $|B_z|<0.1$~kG (called "non-magnetic"). Interestingly, we obtain that the average line strengthening and weakening is almost the same for the magnetic  and the non-magnetic areas. However, the width of the distribution is much larger for the magnetic areas, most magnetic pixels are associated with a line weakening in \mbox{3-D}, but the distribution has a long tail towards great line strengthening. Strong down-flows which mainly take place outside magnetic elements harbor line weakening in \mbox{3-D} NLTE as they have hot surroundings. Furthermore, as mentioned above, in the innermost parts of magnetic elements considerable line strengthening takes place.

\begin{comment}
average bei middle panel  dashed thick:  
1      0.51029312	vergleiche mit Fig extreme samples, Punkt ist dass \mbox{3-D} resp \mbox{1-D} Linie extrem schwach ist
2      0.10515055
FLUXAREA        LONG      = Array[11268]
NOFLUXAREA      LONG      = Array[16814]
UPFLOWS         LONG      = Array[7119]
DOWNFLOWS       LONG      = Array[12127]
ATMOSPHERE      LONG      = Array[82944]
\end{comment}

\subsection{Line formation in a realistic flux tube}\label{sec:mhdfe23_results_FT}
%%%%%%%%%%%%%%%%%%%%%%%%%%%%%%%%%%%%%%%%%%%%%%%%%%%%%%%%%%%%%%%%%%%%%%%%
In the previous section we found lines formed in \mbox{3-D} NLTE to be stronger than their \mbox{1-D} NLTE counterparts in the centers of magnetic elements (i.e. $\delta E^{1D-3D}<0$, in contradiction to previous results from idealized models. In order to better understand this surprising result and to compare it with the results of \mbox{Paper I}, we selected a sample cut through a magnetic element that is FT like. The cut runs diagonally and is centered on $x \approx 1.8, y \approx 3.3$~Mm as indicated in Fig.~\ref{fig:tv_overview_Icont}. 

\subsubsection{Line profiles}\label{sec:mhdfe23_results_profiles}
%%%%%%%%%%%%%%%%%%%%%%%%%%%%%%%%%%%%%%%%%%%%%%%%%%%%%%%%%%%%%%%%%%%%%%%%
\begin{figure*}
\center{ \resizebox{0.80\width}{!}{\includegraphics{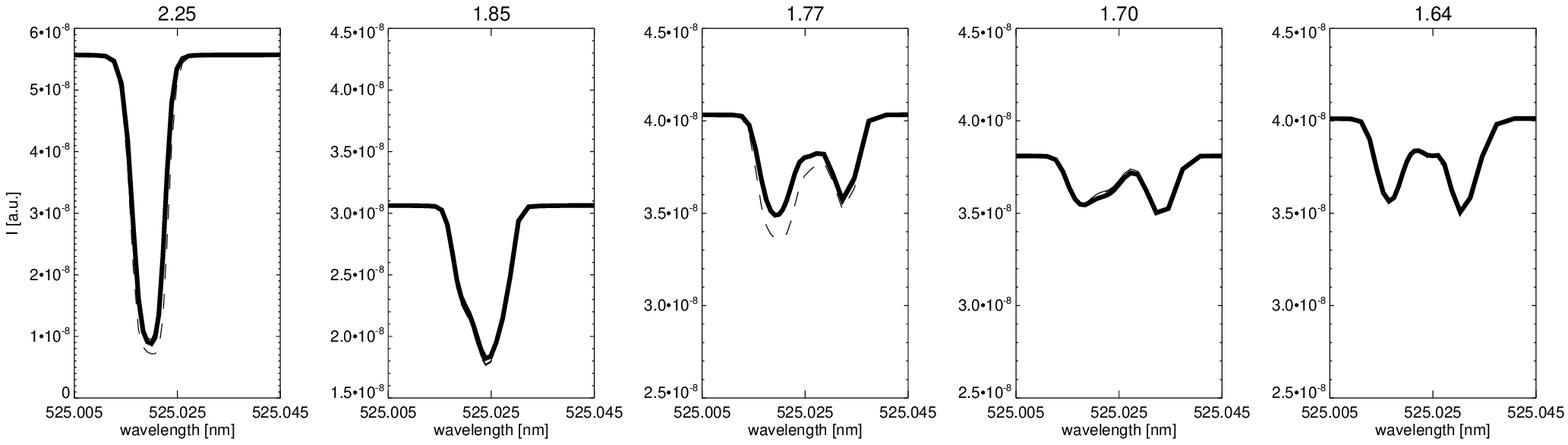}} }
\center{ \resizebox{0.80\width}{!}{\includegraphics{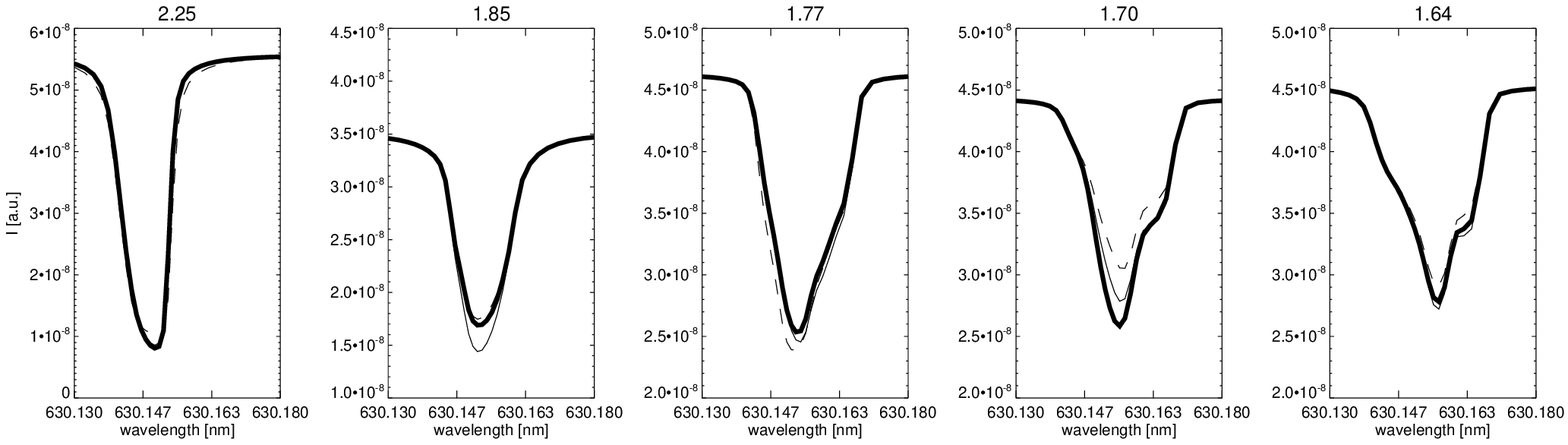}} }
 \caption{ 
Sample profiles at selected positions along the cut through the MHD FT. The $x$-positions from where the profiles originate are indicated above each panel. Line styles: thick solid: \mbox{3-D} NLTE; thin solid: \mbox{1-D} NLTE; dashed: LTE. Note the different $y$-axis ranges used in the panels.
} 
 \label{fig:sample_profiles}
\end{figure*}

\begin{figure}
\center{ \resizebox{0.5\width}{!}{\includegraphics{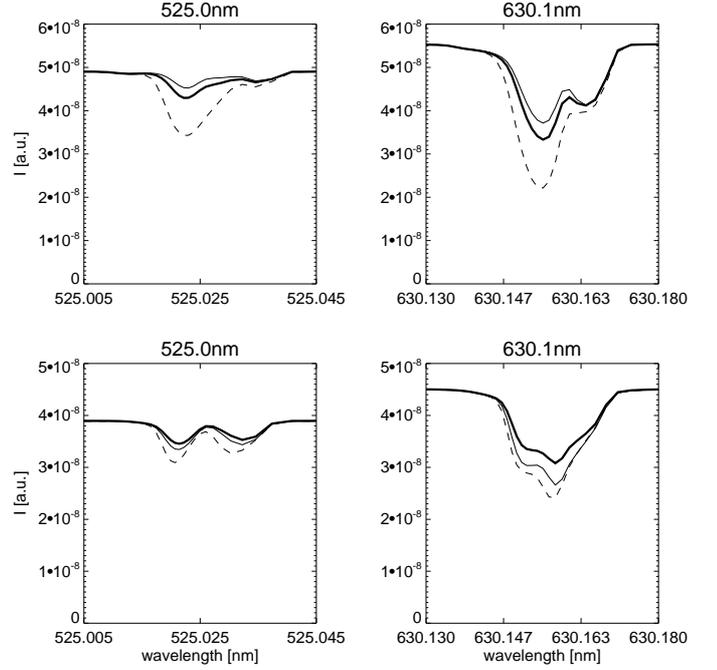}} }
 \caption{ 
 Two extreme examples of the influence of NLTE and horizontal RT on line profiles: left panels: \ion{Fe}{i} $525.02$~nm line; right panels: \ion{Fe}{i} $630.15$~nm line; upper panels: $x \approx 2.15$~Mm, $y \approx 0.4$~Mm; lower panels:  $x \approx 3.61$~Mm, $y=1.50$~Mm. Line styles as in Fig.~\ref{fig:sample_profiles}.  
} 
 \label{fig:extreme_profiles}
\end{figure}

Profiles of \ion{Fe}{i} $525.02$~nm and $630.15$~nm are plotted in Fig.~\ref{fig:sample_profiles} at five spatial locations along the cut through the FT used in the previous section (see Fig.~\ref{fig:tv_overview_Icont}). The profiles between these positions typically show a continuous, slow change towards the next plotted profile. At position $x=2.25$~Mm, in the middle of a granule, the profiles are deep, with a strong continuum. Small differences on the order of $1$\% to $3$\% (LTE) are found between the line depths of the three profiles at this spatial location, the \mbox{1-D} NLTE line being slightly weaker than the \mbox{3-D} NLTE line, as expected from \mbox{Paper II}. At position $x=1.85$~Mm near the boundary of the FT, the line intensities as well as the continuum are much weaker and, owing to the Zeeman effect, the line is broader. At this position, we find $\delta E^{1D-3D}>0$, i.e. line weakening, in agreement with idealized FT. The \ion{Fe}{i} $525.02$~nm line, however, is so weak and formed so deep that only minor absolute differences remain. Note the different $y$-axis ranges. At $x=1.77$~Mm, in the inner FT, the lines are even broader and the \ion{Fe}{i} $525.02$~nm line is completely split by the Zeeman effect. Here we find $\delta E^{1D-3D} \approx 0$. In the core of the FT, at $x=1.70$~Mm, $\delta E^{1D-3D}$ is negative for the $630.15$~nm line, and the $525.02$~nm line is so weak that no apparent difference may be seen. At $x=1.64$~Mm we find $\delta E^{1D-3D} \approx 0$  on the other side of the FT. At spatial locations between $x=1.64$~Mm and the outer boundary of the FT $\delta E^{1D-3D}$ is again positive, representing the "normally expected" case of line weakening. 

The profiles along our cut represent \emph{standard} cases. The differences between the individual profiles are mostly small. There are also locations within the analyzed snapshot with much larger differences. Fig.~\ref{fig:extreme_profiles} presents two such cases where $|\delta E^{1D-3D}|$ reaches up to several $10$\%. Although the differences in the profiles are much larger, the two locations still represent analogous situations: The profiles in the upper panels originate from the center of a FT, whereas those shown in the lower panels emerge from a FT boundary. Note, that the LTE profiles are much stronger in both cases, i.e. the inclusion of horizontal RT may either increase or decrease the differences to LTE values.
% delta EW (1D-3D)/3D = 35%  lower panels: 24%

\subsubsection{Atmospheric conditions in the sample flux tube-like structure}\label{sec:mhdfe23_results_atmosFT}
%%%%%%%%%%%%%%%%%%%%%%%%%%%%%%%%%%%%%%%%%%%%%%%%%%%%%%%%%%%%%%%%%%%%%%%%
\begin{figure}
\center{ \resizebox{0.75\width}{!}{\includegraphics{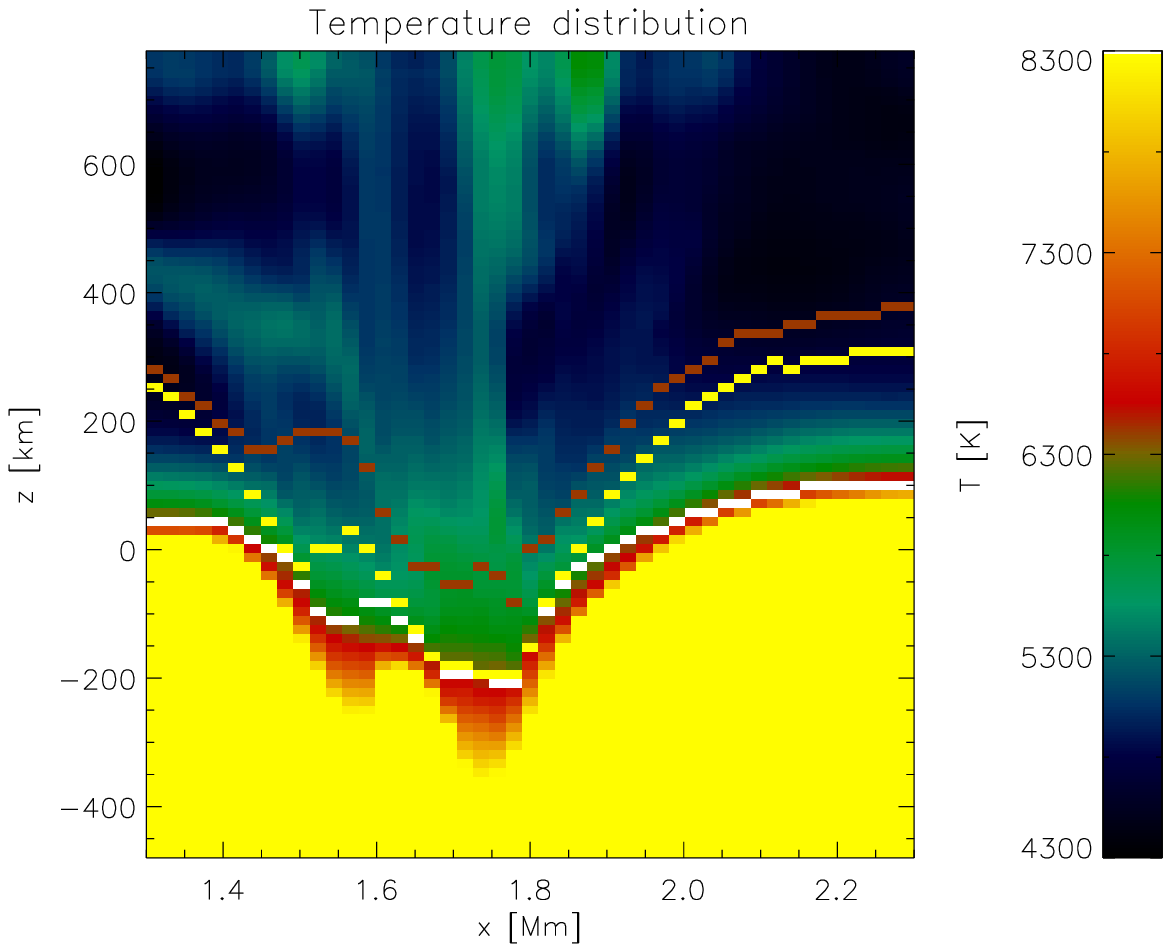}} }
\center{ \resizebox{0.75\width}{!}{\includegraphics{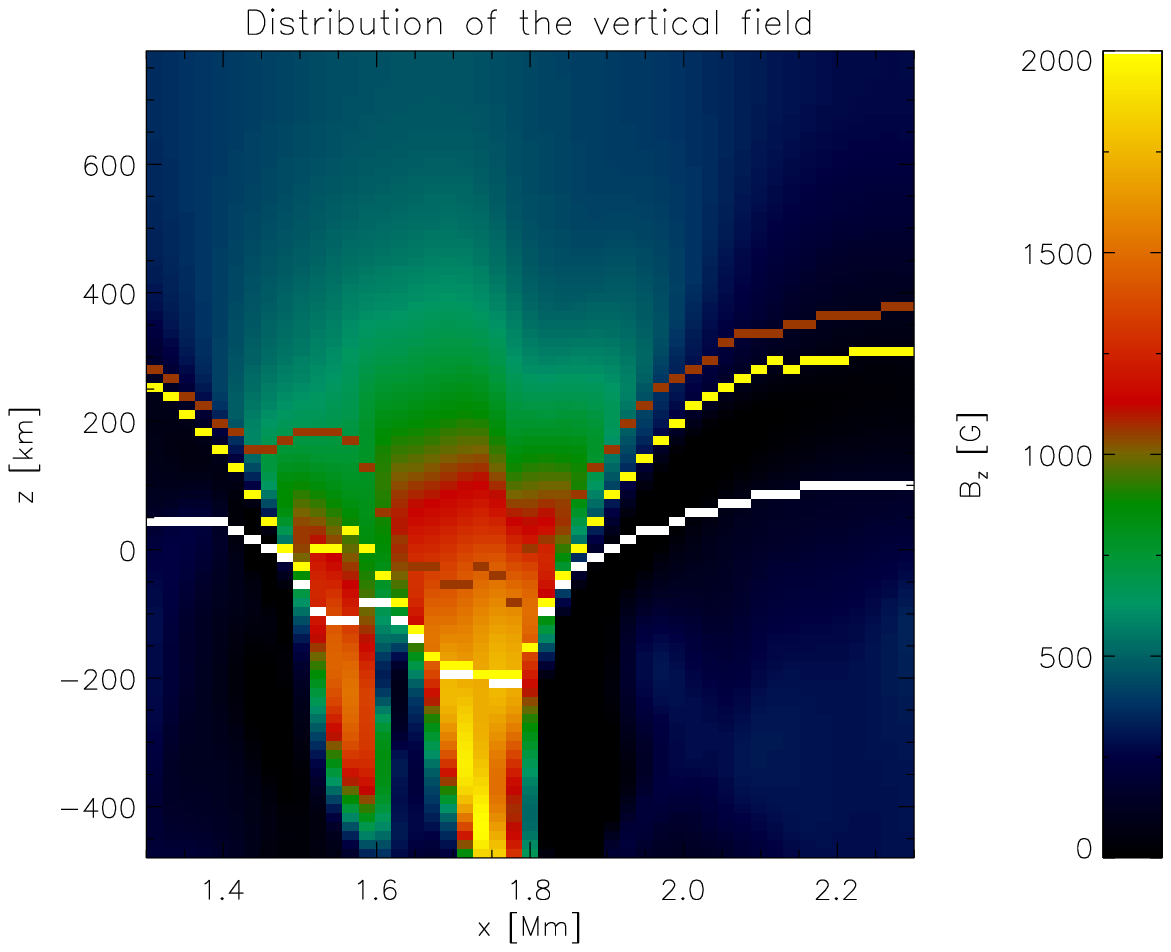}} }
 \caption{ 
 Temperature distribution (upper panel) and vertical magnetic field $B_z$ (lower panel) along the cut through the FT indicated in Fig.~\ref{fig:tv_overview_Icont} (diagonal line). The colored dots indicate several $\tau=1$ levels (white: continuum near $525$~nm; yellow: $525.02$~nm line core; brown: $630.15$~nm line core). Note that the values of the $x$-axis correspond to those of the x-axis of the simulation and not to the length along the diagonal cut. I.e., the true distance between two points along the (diagonal) cut is longer by a factor $\sqrt{2}$.} 
 \label{fig:cut1_tv_temp}
 \label{fig:cut1_Bz_temp}
\end{figure}

Figure \ref{fig:cut1_tv_temp} represents the temperature profile (upper panel) and the vertical magnetic field strength (lower panel) along the cut. The dotted lines in white indicates the continuum ($\tau_c=1$) level. At most horizontal locations (granules, center of FT/FS) the continuum is formed in the temperature range above $~6300$~K, i.e. the region colored in red. However, it is formed in a cooler layer (green layer with $\approx 6000$~K) close to the borders of the FT where the vertical gradient of the temperature is smaller (compare the thickness of the green and red layers at the different positions, e.g., at $x=1.8$~Mm with $x=2.2$~Mm). This corresponds well with the brightness of the continuum in Fig.~\ref{fig:tv_overview_Icont}. By comparing the two panels of Fig.~\ref{fig:cut1_tv_temp} we clearly see that the $\tau=1$ level starts to curve down well outside the magnetic feature. Only a small part of this change in $\tau=1$ outside the FT is due to the expansion of the FT, which places low density gas above these locations (see Paper I). It is mainly due to the lower temperature in the downflow lanes.

The yellow and brown lines in Fig.~\ref{fig:cut1_tv_temp} denote the $\tau=1$ level of the $525.02$~nm and the $630.15$~nm line cores, respectively. Due to its strong temperature sensitivity, the $525.02$~nm line is formed almost at the same height as the continuum over large parts of the FT thus minimizing the possible influence of NLTE effects in general. The height of its formation in the granules amounts to about $200$~km above $\tau_c=1$, i.e., the Wilson depression in the $525.02$~nm line core is far larger than that in the continuum, whereas it is only about $100$~km larger than that of the continuum in the $630.15$~nm line core.

\begin{figure}
\center{ \resizebox{0.75\width}{!}{\includegraphics{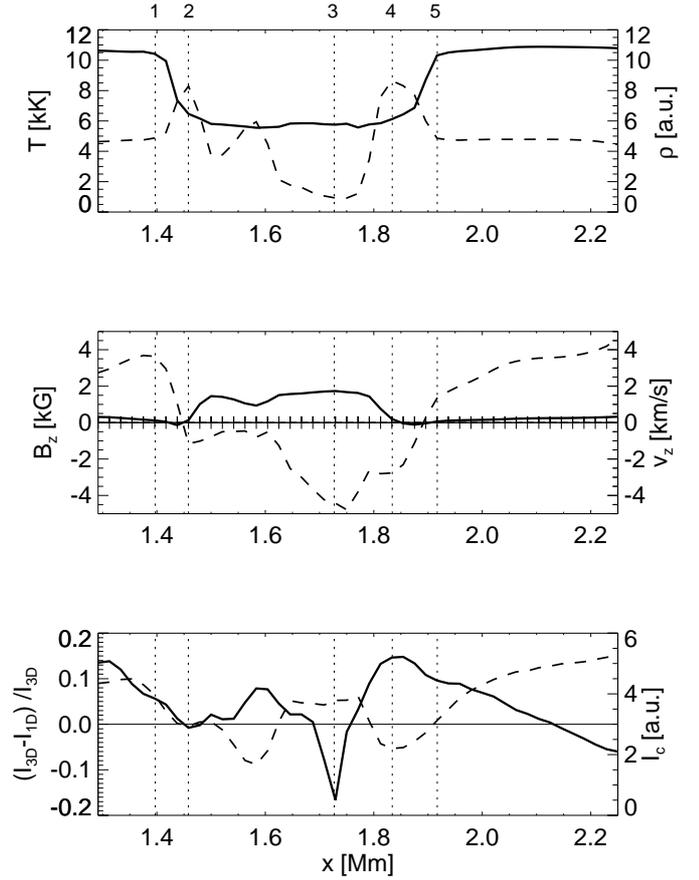}} }
 \caption{ 
Physical quantities at a fixed geometrical height ($z=-42$~km) along the same diagonal cut as in Fig.~\ref{fig:cut1_tv_temp}. This height corresponds roughly to the line core formation height of the $630.15$~nm line in the center of the FT. The axis description of the solid lines is given on the left, that of the dashed lines on the right side. Top panel: Temperature (solid) and density (dashed). Middle panel: Magnetic field (solid) and vertical velocity (dashed, positive values correspond to upflows). The tick marks at the horizontal zero line correspond to the individual voxels in the model atmosphere. Bottom panel: NLTE intensity ratio $(I_{\mbox{3-D}} - I_{\mbox{1-D}}) / I_{\mbox{3-D}}$ (solid) representing \mbox{3-D} NLTE line weakening (positive values) or strengthening (negative values) and  continuum intensity (dashed). Vertical lines (dotted) represent positions (numbered on top of the uppermost panel) along the cut that are discussed in the main text. As in Fig.~\ref{fig:cut1_tv_temp}, the values of the $x$-axis correspond to x-axis of the simulation and not to the length along the diagonal cut.}
\label{fig:cut1_divquant}
\end{figure}

Fig.~\ref{fig:cut1_divquant} presents some quantities along the cut through the FT at a fixed geometrical height ($z=-42$~km). The $630.15$~nm line core is formed at this approximate height in the FT. For better visibility and comparison of the different panels, some of the horizontal $x$-positions that are mentioned in the following are marked with a thin dotted vertical line (numbered on top of the uppermost panel). Several interesting facts can be seen from Fig.~\ref{fig:cut1_divquant}:
\begin{enumerate}
\item[a] top panel: The drop in temperature (thick line) associated with the FT starts at $x=1.40$~Mm (pos.~1) and ends at $x=1.92$~Mm (pos.~5). This drop is seen at a fixed geometric height. There are two boundary regions with enhanced density (dashed line) in the cool part of the FT of the atmosphere.
\item[b]  top and middle panels:   The maximum density is reached at the positions 2 and 4 just outside the boundary of the FT where the magnetic field (thick line) has dropped to nearly zero. 
\item[c] top and middle panel: According to the magnetic field strength the FT diameter at the chosen height amounts to approximately $470$~km (taking into account the $\sqrt 2$ factor because the cut is in diagonal direction). The region with low temperature is considerably larger, i.e., approximately $650$~km.
\item[d] middle panel: The drop in temperature occurs over a small range in $x$ at the boundary between granular up- and down-flows, i.e, at the edges of the granules.
\item[e] bottom panel: The spectral line formed in \mbox{3-D} NLTE is weakened relative to that in \mbox{1-D} NLTE  (thick line, positive values) in most locations along the cut, except in the strongest field part of the FT and in the granule to the right ($x>2.1$~Mm). The strengthening in the granule is expected from the results of \mbox{Paper II}). The granule on the left is small and has a complex structure (see continuum image in Fig.~\ref{fig:tv_overview_Icont}), so that the effect is not so clean.
\item[f] bottom panel: The \mbox{3-D} NLTE line is strengthened in the core of the FT (pos.~3). The temperature is also slightly higher at that location.
\item[g] bottom panel: The continuum intensity (thin line) roughly corresponds inversely to the line weakening/strengthening. The locations of lowest continuum intensity coincide with peaks in the gas density.
\end{enumerate}

According to Fig. 1 in \mbox{Paper I} the continuum in the idealized FT is formed at a depth at which the surrounding walls are at a temperature of $10000$~K. In contrast, the walls in the MHD atmosphere, i.e. the region where the density increases and the field strength decreases, harbor down-flows which are not significantly hotter than the interior of the magnetic element at the same geometric height. This cool gas immediately surrounding the FT has an even higher density than that in the granules at the same geometrical height. The higher temperature reached in granules is located further away from the magnetic feature, behind the humps of high density.

\begin{figure}
\center{ \resizebox{0.75\width}{!}{\includegraphics{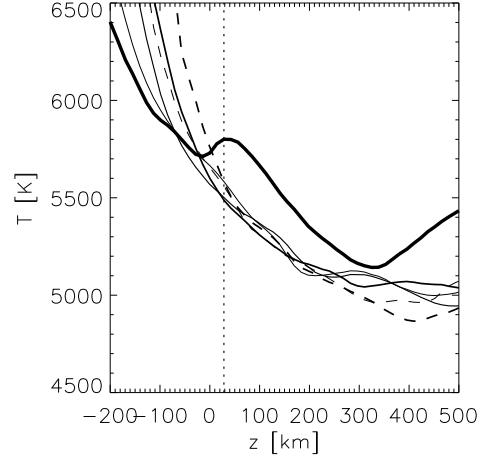}} }
 \caption{ 
Temperature curves as a function of height in the core of the chosen FT where the 3-D NLTE line strengthening is strongest (thick solid line) and at positions in the surroundings of the FT. Other solid lines: temperature curves averaged over the four points in positive and negative $x$ and $y$ direction at a given distance from the FT center. The thickness of the line increases with distance from the FT core. From thinnest to thicker lines (solid): $84$~km, $146$~km, and, $209$~km. Dashed lines: the same as for the solid ones but for the four points in diagonal direction. The two lines lie $118$~km (thinner line) and $177$~km (thicker line) away.
The vertical line indicates the approximate height of formation of the $630.15$~nm line core in the core of the FT, where the continuum is formed slightly below $z=-200$~km.}
\label{fig:cut1_FT_Tenv}
\end{figure}

The \mbox{3-D} NLTE line strengthening in the core of the FT is a consequence of the walls being not as hot as in the idealized FT case. Contributions from the very hot granular gas (with $T > 7000$ K) are absorbed before they reach the inner part of the FT. Furthermore, the immediate surrounding of the FT core is cooler on average than the core itself. This is illustrated in Fig.~\ref{fig:cut1_FT_Tenv}, where the temperature curve at the FT core (bold line) and the run of the average temperature of the surroundings at different distances and directions (thin lines, see figure caption) is given. The temperature in the FT core at the height of formation of the $630.15$~nm line core (vertical line) is several $100$~K higher than that in its immediate surrounding. The run of the temperature much swallower than in the atmosphere outside the FT. These two facts lead to the situation that an atom at the center of the FT senses less UV irradiation than an atom closer to the boundaries.

\subsection{Conditions in a weak flux sheet}\label{sec:mhdfe23_results_FS}
%%%%%%%%%%%%%%%%%%%%%%%%%%%%%%%%%%%%%%%%%%%%%%%%%%%%%%%%%%%%%%%%%%%%%%%%
\begin{figure*}
\center{ \resizebox{0.85\width}{!}{\includegraphics{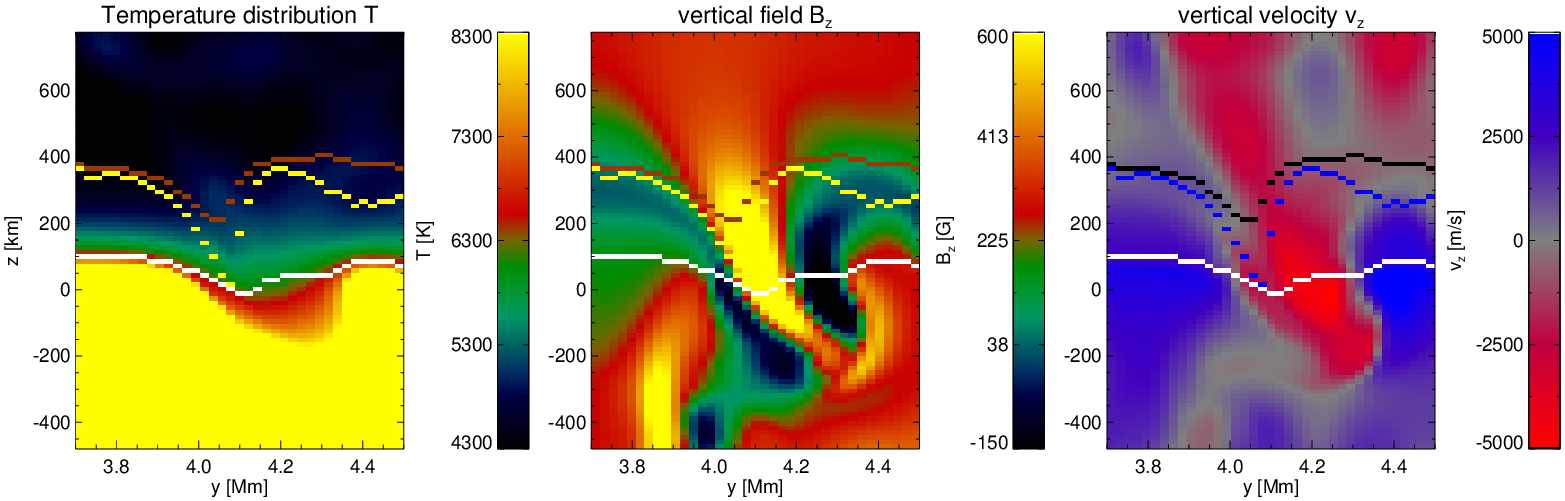}} }
 \caption{ 
 Temperature profile (left panel), vertical magnetic field $B_z$ (middle panel), and vertical velocity $v_z$ (right panel) in a vertical cut through a sample FS. The position of the cut is marked in the continuum image of Fig.~\ref{fig:tv_overview_Icont} (vertical cut near the right upper edge). Additionally, the formation heights of the same spectral line cores as in Fig.~\ref{fig:cut1_tv_temp} are given: Left and middle panel: Continuum (white line), $525.02$~nm line core (yellow line), $630.15$~nm line core (brown line). In the right panel the corresponding colors are chosen differently: white (continuum), blue ($525.02$~nm line core), and, black ($630.15$~nm line core).
} 
 \label{fig:cut2_tv}
\end{figure*}

\begin{figure}
\center{ \resizebox{0.70\width}{!}{\includegraphics{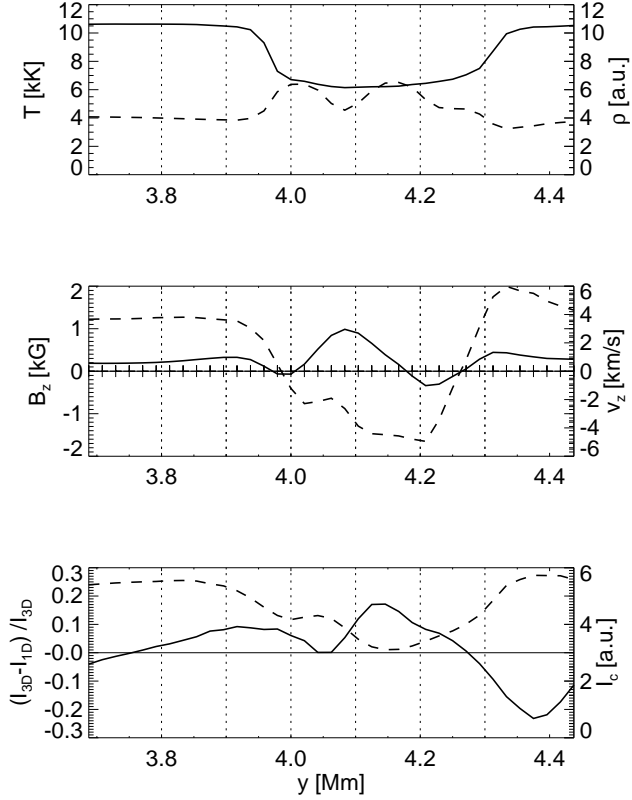}} }
 \caption{ The same quantities as in Fig.~\ref{fig:cut1_divquant} along the cut through the FS at geometrical height $z=0$~km. 
} 
 \label{fig:cut2_divquant}
\end{figure}

The MHD snapshot used in this work, shows two typical forms of FS. The two forms can be distinguished in the continuum image of Fig.~\ref{fig:tv_overview_Icont}. Many intergranular lanes are dark and have a central brightening. They show a similar behavior like that of the FT discussed in the previous section. The second form of intergranular lanes, which will be discussed here, shows no central brightening in the continuum. Figure \ref{fig:cut2_tv} presents several atmospheric quantities along a cut through such a FS, which, in this case, is tilted towards a neighboring granule, a situation quite common in the analyzed simulation snapshot. The main reason for the difference in behavior is the lower field strength in this FS \citep[see][]{TritschlerUitenbroek2006,riethmuelleretal2014,riethmuellersolanki2015}. The temperature in the FS (left panel) is quite similar to that in the FT. However, the magnitude of the temperature fluctuations is smaller, as expected from the weaker magnetic field (see middle panel). In the left panel the continuum and line formation height is indicated, as in Fig.~\ref{fig:cut1_tv_temp}. The continuum is formed in two different layers. In the granules, the formation height is in the temperature range above $6300$~K (red parts), whereas it is within layers of approximately $5600$~K (green layer) in the down-flowing parts of the atmosphere closer to and in the FS. 

The $525.02$~nm line displays again a much larger Wilson depression than the continuum. The Wilson depression in the core of \ion{Fe}{i} $630.15$~nm is of intermediate magnitude. Owing to the tilt of the FS, the location of deepest origin follows the location of the strongest field at the height of formation, shifting from the left to the right in the left panel of Fig.~\ref{fig:cut2_divquant} when going from the  $630.15$~nm towards the more weakened $525.02$~nm line and the continuum.

From the middle panel of Fig.~\ref{fig:cut2_tv}, we see that $B_z$ is of the order of several hG in the FS. The maximum value amounts to approximately $1$ kG. Therefore, the corresponding Wilson depression is much smaller than that in the FT discussed in Sect.~\ref{sec:mhdfe23_results_FT}. The FS is surrounded by two regions of reversed $B_z$ of weaker field strength ($150$ G, dark blue). The strong magnetic field in the FS does not extend as deep as that of the FT. The strong turbulence of the granular atmosphere deforms the relatively weak field and breaks up the sheet in deeper layers. This is typical of many FS-like structures and the one shown here is representative in this respect. 

Figure \ref{fig:cut2_divquant} presents the same quantities at height $z=0$~km as Fig.~\ref{fig:cut1_divquant}, but now along the cut through the FS. The situation is very similar: There are again two cool, density enhanced walls on both sides of the FS. The density decreases as the magnetic field increases, but interestingly remains above the value in the granules on either side. The width of the magnetic FS is much smaller than that of the region of cool gas, i.e. the walls are, at a given height, mainly at the same temperature as the interior of the FS. The strongest down-flow is found at the boundary of the FS where the magnetic field direction is reversed (probably due to the strong down-flow pulling the weak magnetic field downwards). Such reversed polarity fields surrounding magnetic elements were detected observationally by \citet{zayeretal1989,narayan2011,buehler2014,buehleretal2015}.

From the bottom panel of Fig.~\ref{fig:cut2_divquant} we see also that line weakening ($\delta E^{1D-3D}>0$) takes place  over the full width of the FS and in its immediate surroundings. However, $\delta E^{1D-3D}$ dips to values very close to zero near the middle of the FS. If one compares with Fig.~\ref{fig:cut1_Bz_temp} and \ref{fig:tv_ratio_Aeqw} one finds that the line weakening is strongest in the intergranular lanes but outside magnetic features. It is smallest in the magnetic features, irrespective of whether the FS core is bright or dark in the continuum (Fig.~\ref{fig:tv_overview_Icont}).  There are only a few exceptions to that rule. E.g. the lane near ($x=2.6$~Mm, $y=2.2$~Mm), for which the line weakening is most pronounced in the middle of the FS. However, this FS has $B<=500$~G at all heights in the photosphere. Such small $B$ values lead to a minute Wilson depression so that $B$ does not influence the horizontal RT in a noticeable manner.

In summary, we find that flux sheets display line weakening due to horizontal irradiation in magnetic features as described by \citet{stenholmstenflo1977}, although in a mild form, due to the presence of dense cool material on both sides of the FS. The inverse effect found in the centers of the FT, is less pronounced or not present owing to the lower field strength.

\section{Discussion}\label{sec:mhdfe23_discussion}
%%%%%%%%%%%%%%%%%%%%%%%%%%%%%%%%%%%%%%%%%%%%%%%%%%%%%%%%%%%%%%%%%%%%%%%%
In their pioneering investigations of \mbox{3-D} radiative transfer in heavily idealized flux tubes, \citet{stenholmstenflo1977,stenholmstenflo1978} found considerable net weakening of \ion{Fe}{i} lines owing to the strong horizontal UV-irradiation from the hot walls of the strongly evacuated flux tubes \citep{spruit1976}. This result was corroborated by the work of \citet{brulsvdluehe2001} in idealized \mbox{2-D} and, in \mbox{Paper I}, in idealized \mbox{3-D} models of FS and FT, although the effect was clearly weaker in these newer studies because of the more realistic UV opacity used. In \mbox{Paper II}, using a snapshot of a HD simulation as the input atmosphere, we found line strengthening, owing to  below-average UV irradiation from colder environments, to be as frequent as line weakening due to excess UV irradiation from hot surroundings. To investigate the situation in a plage region, we here applied our full \mbox{3-D} NLTE RT calculation to a snapshot of an MHD simulation carried out with the MURAM code \citep{voegleretal2005} with an initial homogeneous vertical field of $400$ Gauss. Naively one would expect the line weakening to dominate over the line strengthening in such an MHD snapshot because of the systematic evacuation of magnetic elements whose interiors are expected to be considerably cooler than their hot surrounding walls at equal geometric height.

\subsection{MHD simulations versus idealized models}\label{sec:mhdfe23_discussion_cartoon}
%%%%%%%%%%%%%%%%%%%%%%%%%%%%%%%%%%%%%%%%%%%%%%%%%%%%%%%%%%%%%%%%%%%%%%%%
The situation in a more realistic plage atmosphere is far more complex than the simple, heavily idealized models used in the past. This leads to the surprising result that horizontal RT not just weakens lines in the magnetic features, in some locations (mainly in the centers of magnetic elements), it actually strengthens the lines.

In idealized models, such as the thin-tube model, the wall may be defined in a simple way as the region where the transition from the outer, hot, dense and field-free to the inner, cool, thin and strongly magnetized atmosphere takes place. In \mbox{Paper I} we defined the wall to be the region where the density decreases from 90\% to 10\% of the density outside the flux-tube (i.e. of the non-magnetic quiet-Sun atmosphere at the same geometrical height), which also coincides with the magnetic field strength increase at the FT boundary. In the thin-tube approximation the temperature also decreases smoothly across the wall from very high values ($10000$~K and more) in the outer  atmosphere towards low values ($5000$~K and less) in the inner parts of the FT/FS at equal geometrical height. Therefore, the inner parts of the magnetic element see hot walls.

In the MHD snapshot the walls, i.e. the high-density regions from which the horizontal radiation reaching the core of the magnetic element originates, are generally cool, due to the presence of strong, cool down-flows around the magnetic elements. The drop in density coincides with the increase of the magnetic field strength, but the temperature drop lies further away because the down-flows extend clearly beyond the magnetic field. Therefore, the magnetic element is radiatively almost completely insulated from the outer hot atmosphere corresponding to granular upflows. Only little or no UV-irradiation from the hot granular gas reaches the core of sufficiently intense and large magnetic elements. This leads to a situation with strong UV irradiation from below, but a sub-average one from the side in the cores of magnetic elements. Consequently, the \mbox{1-D} NLTE lines are weaker than their \mbox{3-D} NLTE counterparts. That the UV irradiation from below is strong may also be seen from the EW ratios of the lines calculated in LTE and \mbox{1-D} NLTE (see Fig.~\ref{fig:tv_ratio_Aeqw}). The \mbox{1-D} NLTE lines are much weaker than those calculated in LTE as the latter ignore over-ionization due to the UV irradiation. Only near the boundary of the elements is the irradiation from the hot granules from the sides still effective and leads to the expected line weakening. 

Because of the high density in the down-flows the continuum is formed at greater heights, i.e., lower temperatures in these walls of the magnetic element. This may also be seen in the continuum image of Fig.~\ref{fig:tv_overview_Icont}, where the granules and the central part of FT/FS are brighter than the intervening lanes. 
 
In weaker magnetic elements, i.e. mainly in FS along intergranular lanes, the line weakening is still the major effect, similar to the idealized models. Although there are cool and dense down-flows on two sides of the FS, the UV irradiation from the hot granule into the element still prevails. Towards the center, the under-population of \ion{Fe}{i} levels due to horizontal UV irradiation is typically strongly suppressed, but a clear strengthening of the lines due to horizontal transfer is found only in a minority of cases.

\subsection{MHD versus HD model}\label{sec:mhdfe23_discussion_HD}
%%%%%%%%%%%%%%%%%%%%%%%%%%%%%%%%%%%%%%%%%%%%%%%%%%%%%%%%%%%%%%%%%%%%%%%%
The plage model used in this work gives similar spatial averages for the strengths of the investigated lines as those we obtained in \mbox{Paper II} for the HD model with $B=0$. Both, line weakening and strengthening, are present in each snapshot and, when averaged spatially over the whole snapshot, the differences between lines calculated in \mbox{1-D} NLTE and those calculated in \mbox{3-D} NLTE are small. On average, for every cool location in the atmosphere with a hot environment we also find a hot location with cool environment. Thus, the global results for the atmospheres with a \mbox{$400$ G} average field and with no field are very similar, although they do not have the same origin when examined in detail.

Furthermore, in both models large differences may occur locally, i.e. if spatially resolved spectra are considered. Local differences of several ten percent are found, those in the MHD snapshot being slightly stronger since the variation of atmospheric parameters, especially the density, is larger in the MHD model. The main differences between the two simulations occur in the intergranular lanes and their intersections. 

The magnetic field in the MHD model impedes the intergranular down-flows, which are therefore partially displaced towards the granules. The magnetic elements themselves are located in the middle of these cool and dense down-flows that separate the hot granular atmosphere from the magnetic element. This diminishes the influence of the hot granular atmosphere on the interior of the magnetic element. In the core of a strong-field element it can even overcompensate for the systematic evacuation of the element. In the HD model, the transition from the granular to the intergranular atmosphere is smoother and resembles more the walls found in the classical thin-FT model. Therefore, in the HD model, the weakening of the lines computed in \mbox{3-D} NLTE spans the whole intergranular lanes, though in a milder form, as the density difference between granules and intergranular lanes is not so large owing to the absence of a magnetic field.

\subsection{Are the FT/FS walls resolved?}\label{sec:mhdfe23_discussion_resolution}
%%%%%%%%%%%%%%%%%%%%%%%%%%%%%%%%%%%%%%%%%%%%%%%%%%%%%%%%%%%%%%%%%%%%%%%%
In \mbox{Paper I} we set the wall thickness of the mainly investigated FT model to be $20$~km. Although, the situation in the MHD FT investigated here is more complex, we can still take the drop from $90$ to $10$ percent in density at the approximate geometrical height of $\tau_c=1$ in the close granular environment as a measure for the wall thickness. For a FT with a horizontal extension of $\approx 500$~km and for a FS of similar width, the density drop typically takes place over two to four voxels, i.e. over a distance of approximately $50$ to $100$~km (see Figs.~\ref{fig:cut1_divquant} and ~\ref{fig:cut2_divquant}). Whether the relatively crude horizontal resolution of $21$~km in the original MHD simulation manages to resolve the walls of magnetic elements is unclear. As the thickness of and the temperature distribution in the wall is directly responsible for the strength of the effects of horizontal RT (see the results of the tests carried out in \mbox{Paper I}), it could well be that a better resolved snapshot may lead to a different quantitative result. We plan to test the influence of resolution on the results of \mbox{3-D} RT in MHD simulations in an upcoming investigation.

\section{Conclusions}\label{sec:mhdfe23_conclusions}
%%%%%%%%%%%%%%%%%%%%%%%%%%%%%%%%%%%%%%%%%%%%%%%%%%%%%%%%%%%%%%%%%%%%%%%%

In this work, we continued our investigation of the influence of horizontal radiative transfer (RT) on diagnostically important \ion{Fe}{i} lines by calculating the line formation in a snapshot of a \mbox{3-D} radiation magneto-hydrodynamic (MHD) simulation performed with the MURAM code \citep{voegleretal2005}. The chosen snapshot represents a strong plage region (the simulation started with an initial vertical homogeneous field of $400$~Gauss). By comparing profiles computed in LTE, \mbox{1-D} NLTE and true \mbox{3-D} NLTE, we are able to isolate and quantify the effects (line weakening by UV irradiation from hot environments) of NLTE and horizontal RT as described for the first time by \citet{stenholmstenflo1977} within a simple and highly idealized flux tube (FT) geometry. These results were confirmed within 2-D \citep{brulsvdluehe2001} and \mbox{3-D} (\mbox{Paper I}) idealized models. We know from \mbox{Paper II}, that in pure hydrodynamic simulations, the opposite effect, i.e. a strengthening of the \mbox{3-D} NLTE lines takes place almost equally often, mainly inside granules. However, for a plage region one would naively expect a clear dominance of the line weakening effect due to the strong UV irradiation from hot walls into the evacuated magnetic elements. 

The surprising result of our computations is that the expected line weakening in the \mbox{3-D} NLTE lines only takes place in the outer parts of a FT. At the center of the FT, the \mbox{3-D} NLTE lines are either unaffected by the horizontal radiation inflow or are even strengthened. The reason for this unexpected behavior lies in the fact, that the FT in the MHD model are surrounded by dense and relatively cool down-flowing gas. This gas forms an isolating buffer between the magnetic element and the hot atmosphere of the neighboring granules. The UV-irradiation from these hot parts of the photosphere, which dominates the behavior in thin-tube models, is therefore strongly suppressed and has a noteworthy effect near the boundaries of the FT only. Even there, it is reduced due to the absorption by the intervening cool and dense down-flowing gas. The down-flows are usually not hotter than the inner part of the FT at equal geometrical height. The continuum in these dense down-flows is formed in higher and therefore cooler layers than in the granules or the innermost part of the FT. This can be clearly seen from the continuum image in Fig.~\ref{fig:tv_overview_Icont}. As the (UV-) continuum irradiation from these surrounding down-flows is often lower than from directly below, the \mbox{3-D} NLTE line is stronger than its \mbox{1-D} NLTE counterpart which does not feel these "colder" contributions, from the side.

In the case of flux sheets (FS) along the intergranular lanes, the situation is mostly similar. The magnetic field strength in the thinner FS is, however, often smaller ($\approx 1$~kG in a weak FS) and, therefore, the effects weaker. Line weakening due to horizontal irradiation takes place at the edges of the FS and in the surrounding dense lanes. The opposite effect in the center of the FS is weaker than for the stronger FT or only present in the form of a suppressed line weakening. 

In paper I we showed that the thickness of the down-flows and the wall, i.e. the region where the density drops and the magnetic field increases when moving into the magnetic element, plays a central role for the qualitative result. The horizontal resolution in the employed MHD atmosphere is $21$~km per voxel, i.e. much lower than that in the simplified models used in \mbox{Paper I}. In a future investigation, we plan to consider also simulation snapshots with a finer grid.

The significantly different behavior between the idealized thin-tube model analyzed in Paper I and the MHD simulations considered here introduces a caveat for the uncritical use of the thin-tube model \citep{defouw1976}, which has been widely and very successfully used. Besides being invaluable for studies of the thermal structure of \citep[][]{spruit1976}, MHD wave propagation in \citep[e.g.][]{robertswebb1979}, or, instabilities of \citep[e.g.][]{webbroberts1978,robertswebb1978,spruit1979} magnetic elements, the thin-tube has turned out to describe a number of observations of polarized spectra rather well \citep[e.g.,][]{zayeretal1989,brulssolanki1995}. Our results indicate that the thin-tube approximation cannot capture the complexity of a magnetic element, so that at least in some contexts it should not be employed. However, the thin-tube approximation provides a very good approximation of the magnetic structure of magnetic elements in \mbox{3-D} MHD simulations \citep{yelleschaoucheetal2009}, at least if the second order thin-tube model is used \citep[e.g.][ and later publications of these authors]{pneuman1986,ferriz-masetal1989}. This implies that it is mainly the structure of the external atmosphere which is responsible for the difference between the present results and those of Paper I.

%%%%%%%%%%%%%%%%%%%%%%%%%%%%%%%%%%%%%%%%%%%%%%%%%%%%%%%%%%%%%%%%%%%%%%%%
\begin{acknowledgements} 
%%%%%%%%%%%%%%%%%%%%%%%%%%%%%%%%%%%%%%%%%%%%%%%%%%%%%%%%%%%%%%%%%%%%%%%%
We are grateful to Manfred Sch\"ussler and Robert Cameron for providing the data cubes resulting from the MURAM simulation run, as well as to Han Uitenbroek for making his excellent full Stokes NLTE radiative transfer code available. Jo Bruls has provided the iron model atom. The authors are grateful to Michael Meyer for providing office space and enabling access to the vast computational resources at ETH. R.~H.~appreciates the flexibility of Prof.~Dr.~Norbert Dillier concerning the working hours at the University Hospital of Zurich. This work has been partly supported by the BK21 plus program through the National Research Foundation (NRF) funded by the Ministry of Education of Korea. 
And we kindly thank the referee for his help to improve the manuscript. 
\end{acknowledgements} 
%%%%%%%%%%%%%%%%%%%%%%%%%%%%%%%%%%%%%%%%%%%%%%%%%%%%%%%%%%%%%%%%%%%%%%%%

%%%%%%%%%%%%%%%%%%%%%%%%%%%%%%%%%%%%%%%%%%%%%%%%%%%%%%%%%%%%%%%%%%%%%%%%

\bibliographystyle{aa}
\bibliography{journals,holzreuter}

\begin{thebibliography}{42}
\expandafter\ifx\csname natexlab\endcsname\relax\def\natexlab#1{#1}\fi

\bibitem[{{Athay} \& {Lites}(1972)}]{athaylites1972}
{Athay}, R.~G. \& {Lites}, B.~W. 1972, \apj, 176, 809

\bibitem[{{Auer}(2003)}]{auer2003}
{Auer}, L. 2003, in Astronomical Society of the Pacific Conference Series, Vol.
  288, Stellar Atmosphere Modeling, ed. I.~{Hubeny}, D.~{Mihalas}, \&
  K.~{Werner}, 3

\bibitem[{{Boyarchuk} {et~al.}(1985){Boyarchuk}, {Lyubimkov}, \&
  {Sakhibullin}}]{boyarchuketal1985}
{Boyarchuk}, A.~A., {Lyubimkov}, L.~S., \& {Sakhibullin}, N.~A. 1985,
  Astrophysics, 22, 203

\bibitem[{{Bruls} {et~al.}(1992){Bruls}, {Rutten}, \&
  {Shchukina}}]{brulsetal1992}
{Bruls}, J. H. M.~J., {Rutten}, R.~J., \& {Shchukina}, N.~G. 1992, A\&A, 265,
  237

\bibitem[{{Bruls} \& {Solanki}(1995)}]{brulssolanki1995}
{Bruls}, J.~H.~M.~J. \& {Solanki}, S.~K. 1995, \aap, 293, 240

\bibitem[{{Bruls} \& {von der L{\"u}he}(2001)}]{brulsvdluehe2001}
{Bruls}, J.~H.~M.~J. \& {von der L{\"u}he}, O. 2001, \aap, 366, 281

\bibitem[{{B{\"u}hler}(2014)}]{buehler2014}
{B{\"u}hler}, D. 2014, PhD thesis, University of G{\"o}ttingen, G{\"o}ttingen,
  Germany

\bibitem[{{B{\"u}hler} {et~al.}(2015){B{\"u}hler}, {Lagg}, {Solanki}, \& {van
  Noort}}]{buehleretal2015}
{B{\"u}hler}, D., {Lagg}, A., {Solanki}, S.~K., \& {van Noort}, M. 2015, \aap,
  576, A27

\bibitem[{{Defouw}(1976)}]{defouw1976}
{Defouw}, R.~J. 1976, \apj, 209, 266

\bibitem[{{Ferriz-Mas} {et~al.}(1989){Ferriz-Mas}, {Sch{\"u}ssler}, \&
  {Anton}}]{ferriz-masetal1989}
{Ferriz-Mas}, A., {Sch{\"u}ssler}, M., \& {Anton}, V. 1989, \aap, 210, 425

\bibitem[{{Holzreuter} \& {Solanki}(2012)}]{holzreutersolanki2012}
{Holzreuter}, R. \& {Solanki}, S.~K. 2012, \aap, 547, A46, (Paper I)

\bibitem[{{Holzreuter} \& {Solanki}(2013)}]{holzreutersolanki2013}
{Holzreuter}, R. \& {Solanki}, S.~K. 2013, A\&A, 558, A20, (Paper II)

\bibitem[{{Kiselman} \& {Nordlund}(1995)}]{KiselmanNordlund1995}
{Kiselman}, D. \& {Nordlund}, A. 1995, \aap, 302, 578

\bibitem[{{Lagg} {et~al.}(2010){Lagg}, {Solanki}, {Riethm{\"u}ller},
  {Mart{\'{\i}}nez Pillet}, {Sch{\"u}ssler}, {Hirzberger}, {Feller}, {Borrero},
  {Schmidt}, {del Toro Iniesta}, {Bonet}, {Barthol}, {Berkefeld}, {Domingo},
  {Gandorfer}, {Kn{\"o}lker}, \& {Title}}]{laggetal2010}
{Lagg}, A., {Solanki}, S.~K., {Riethm{\"u}ller}, T.~L., {et~al.} 2010, \apjl,
  723, L164

\bibitem[{{Leenaarts}(2010)}]{leenaarts2010}
{Leenaarts}, J. 2010, \memsai, 81, 576

\bibitem[{{Leenaarts} {et~al.}(2009){Leenaarts}, {Carlsson}, {Hansteen}, \&
  {Rouppe van der Voort}}]{leenaartsetal2009}
{Leenaarts}, J., {Carlsson}, M., {Hansteen}, V., \& {Rouppe van der Voort}, L.
  2009, \apjl, 694, L128

\bibitem[{{Leenaarts} {et~al.}(2012){Leenaarts}, {Carlsson}, \& {Rouppe van der
  Voort}}]{leenaartsetal2012}
{Leenaarts}, J., {Carlsson}, M., \& {Rouppe van der Voort}, L. 2012, \apj, 749,
  136

\bibitem[{{Leenaarts} {et~al.}(2013){Leenaarts}, {Pereira}, {Carlsson},
  {Uitenbroek}, \& {De Pontieu}}]{leenaartsetal2013}
{Leenaarts}, J., {Pereira}, T.~M.~D., {Carlsson}, M., {Uitenbroek}, H., \& {De
  Pontieu}, B. 2013, \apj, 772, 90

\bibitem[{{Mart{\'{\i}}nez Gonz{\'a}lez} {et~al.}(2012){Mart{\'{\i}}nez
  Gonz{\'a}lez}, {Bellot Rubio}, {Solanki}, {Mart{\'{\i}}nez Pillet}, {Del Toro
  Iniesta}, {Barthol}, \& {Schmidt}}]{martinezgonzalezetal2012}
{Mart{\'{\i}}nez Gonz{\'a}lez}, M.~J., {Bellot Rubio}, L.~R., {Solanki}, S.~K.,
  {et~al.} 2012, \apjl, 758, L40

\bibitem[{{Narayan}(2011)}]{narayan2011}
{Narayan}, G. 2011, \aap, 529, A79

\bibitem[{{Pneuman} {et~al.}(1986){Pneuman}, {Solanki}, \&
  {Stenflo}}]{pneuman1986}
{Pneuman}, G.~W., {Solanki}, S.~K., \& {Stenflo}, J.~O. 1986, \aap, 154, 231

\bibitem[{{Rees}(1969)}]{rees1969}
{Rees}, D.~E. 1969, \solphys, 10, 268

\bibitem[{{Riethm{\"u}ller} \& {Solanki}(2015)}]{riethmuellersolanki2015}
{Riethm{\"u}ller}, T.~L. \& {Solanki}, S.~K. 2015, A\&A, submitted

\bibitem[{{Riethm{\"u}ller} {et~al.}(2014){Riethm{\"u}ller}, {Solanki},
  {Berdyugina}, {Sch{\"u}ssler}, {Mart{\'{\i}}nez Pillet}, {Feller},
  {Gandorfer}, \& {Hirzberger}}]{riethmuelleretal2014}
{Riethm{\"u}ller}, T.~L., {Solanki}, S.~K., {Berdyugina}, S.~V., {et~al.} 2014,
  \aap, 568, A13

\bibitem[{{Roberts} \& {Webb}(1978)}]{robertswebb1978}
{Roberts}, B. \& {Webb}, A.~R. 1978, \solphys, 56, 5

\bibitem[{{Roberts} \& {Webb}(1979)}]{robertswebb1979}
{Roberts}, B. \& {Webb}, A.~R. 1979, \solphys, 64, 77

\bibitem[{{Rutten}(1988)}]{rutten1988}
{Rutten}, R.~J. 1988, in Astrophysics and Space Science Library, Vol. 138, IAU
  Colloq. 94: Physics of Formation of Fe II Lines Outside LTE, ed. {R.~Viotti,
  A.~Vittone, \& M.~Friedjung}, 185--210

\bibitem[{{Shapiro} {et~al.}(2010){Shapiro}, {Schmutz}, {Schoell},
  {Haberreiter}, \& {Rozanov}}]{shapiroetal2010}
{Shapiro}, A.~I., {Schmutz}, W., {Schoell}, M., {Haberreiter}, M., \&
  {Rozanov}, E. 2010, \aap, 517, A48

\bibitem[{{Shchukina} \& {Trujillo Bueno}(2001)}]{shchukinatrujillobueno2001}
{Shchukina}, N. \& {Trujillo Bueno}, J. 2001, \apj, 550, 970

\bibitem[{{Solanki} \& {Steenbock}(1988)}]{solankisteenbock1988}
{Solanki}, S.~K. \& {Steenbock}, W. 1988, \aap, 189, 243

\bibitem[{{Spruit}(1976)}]{spruit1976}
{Spruit}, H.~C. 1976, Sol. Phys., 50, 269

\bibitem[{{Spruit}(1979)}]{spruit1979}
{Spruit}, H.~C. 1979, \solphys, 61, 363

\bibitem[{{Steenbock} \& {Holweger}(1984)}]{steenbockholweger1984}
{Steenbock}, W. \& {Holweger}, H. 1984, \aap, 130, 319

\bibitem[{{Stenholm} \& {Stenflo}(1977)}]{stenholmstenflo1977}
{Stenholm}, L.~G. \& {Stenflo}, J.~O. 1977, A\&A, 58, 273

\bibitem[{{Stenholm} \& {Stenflo}(1978)}]{stenholmstenflo1978}
{Stenholm}, L.~G. \& {Stenflo}, J.~O. 1978, A\&A, 67, 33

\bibitem[{{Th{\'e}venin} \& {Idiart}(1999)}]{theveninidiart1999}
{Th{\'e}venin}, F. \& {Idiart}, T.~P. 1999, \apj, 521, 753

\bibitem[{{Tritschler} \& {Uitenbroek}(2006)}]{TritschlerUitenbroek2006}
{Tritschler}, A. \& {Uitenbroek}, H. 2006, \apj, 648, 741

\bibitem[{{Uitenbroek}(2000)}]{uitenbroek2000}
{Uitenbroek}, H. 2000, ApJ, 531, 571

\bibitem[{{V{\"o}gler} {et~al.}(2005){V{\"o}gler}, {Shelyag}, {Sch{\"u}ssler},
  {Cattaneo}, {Emonet}, \& {Linde}}]{voegleretal2005}
{V{\"o}gler}, A., {Shelyag}, S., {Sch{\"u}ssler}, M., {et~al.} 2005, \aap, 429,
  335

\bibitem[{{Webb} \& {Roberts}(1978)}]{webbroberts1978}
{Webb}, A.~R. \& {Roberts}, B. 1978, \solphys, 59, 249

\bibitem[{{Yelles Chaouche} {et~al.}(2009){Yelles Chaouche}, {Solanki}, \&
  {Sch{\"u}ssler}}]{yelleschaoucheetal2009}
{Yelles Chaouche}, L., {Solanki}, S.~K., \& {Sch{\"u}ssler}, M. 2009, A\&A,
  504, 595

\bibitem[{{Zayer} {et~al.}(1989){Zayer}, {Solanki}, \&
  {Stenflo}}]{zayeretal1989}
{Zayer}, I., {Solanki}, S.~K., \& {Stenflo}, J.~O. 1989, \aap, 211, 463

\end{thebibliography}

%%%%%%%%%%%%%%%%%%%%%%%%%%%%%%%%%%%%%%%%%%%%%%%%%%%%%%%%%%%%%%%%%%%%%%%%%

%%%%%%%%%%%%%%%%%%%%%%%%%%%%%%%%%%%%%%%%%%%%%%%%%%%%%%%%%%%%%%%%%%%%%%%%
\end{document}